\documentclass[twocolumn]{aastex63}

\submitjournal{ApJ}

\shorttitle{COSMIC DUST ANALOGUES FROM C, C$_2$, AND C$_2$H$_2$ IN C-RICH CSEs}
\shortauthors{Santoro et al.}

\begin{document}

\title{THE CHEMISTRY OF COSMIC DUST ANALOGUES FROM C, C$_2$, AND C$_2$H$_2$  IN C-RICH CIRCUMSTELLAR ENVELOPES}

\correspondingauthor{Gonzalo Santoro, Jos\'e \'Angel Mart\'i{}n-Gago}
\email{gonzalo.santoro@icmm.csic.es, gago@icmm.csic.es}

\author{Gonzalo Santoro}
\affiliation{Instituto de Ciencia de Materiales de Madrid (ICMM. CSIC). Materials Science Factory. Structure of Nanoscopic Systems Group. c/ Sor Juana In\'es de la Cruz 3, 28049 Cantoblanco, Madrid, Spain.}

\author{Lidia Mart\'\i{}nez}
\affiliation{Instituto de Ciencia de Materiales de Madrid (ICMM. CSIC). Materials Science Factory. Structure of Nanoscopic Systems Group. c/ Sor Juana In\'es de la Cruz 3, 28049 Cantoblanco, Madrid, Spain.}

\author{Koen Lauwaet}
\affiliation{IMDEA Nanociencia. Ciudad Universitaria de Cantoblanco, 28049 Cantoblanco, Madrid, Spain.}

\author{Mario Accolla}
\affiliation{Instituto de Ciencia de Materiales de Madrid (ICMM. CSIC). Materials Science Factory. Structure of Nanoscopic Systems Group. c/ Sor Juana In\'es de la Cruz 3, 28049 Cantoblanco, Madrid, Spain.}

\author{Guillermo Tajuelo-Castilla}
\affiliation{Instituto de Ciencia de Materiales de Madrid (ICMM. CSIC). Materials Science Factory. Structure of Nanoscopic Systems Group. c/ Sor Juana In\'es de la Cruz 3, 28049 Cantoblanco, Madrid, Spain.}

\author{Pablo Merino}
\affiliation{Instituto de Ciencia de Materiales de Madrid (ICMM. CSIC). Materials Science Factory. Structure of Nanoscopic Systems Group. c/ Sor Juana In\'es de la Cruz 3, 28049 Cantoblanco, Madrid, Spain.}
\affiliation{Instituto de F\'\i{}sica Fundamental (IFF. CSIC). Group of Molecular Astrophysics. c/ Serrano 123, 28006 Madrid, Spain}

\author{Jes\'us M. Sobrado}
\affiliation{Centro de Astrobiolog\'i{}a (CAB. INTA-CSIC). Crta- de Torrej\'on a Ajalvir km4, 28850, Torrej\'on de Ardoz, Madrid, Spain.}

\author{Ram\'on J. Pel\'aez}
\affiliation{Instituto de Estructura de la Materia (IEM.CSIC). Molecular Physics Department. c/Serrano 123, 28006 Madrid, Spain.}

\author{V\'i{}ctor J. Herrero}
\affiliation{Instituto de Estructura de la Materia (IEM.CSIC). Molecular Physics Department. c/Serrano 123, 28006 Madrid, Spain.}

\author{Isabel Tanarro}
\affiliation{Instituto de Estructura de la Materia (IEM.CSIC). Molecular Physics Department. c/Serrano 123, 28006 Madrid, Spain.}

\author{\'Alvaro Mayoral}
\affiliation{School of Physical Science and Technology, ShanghaiTech University, 393 Middle Huaxia Road, Pudong, Shanghai, 201210, People’s Republic of China.}

\author{Marcelino Ag\'undez}
\affiliation{Instituto de F\'\i{}sica Fundamental (IFF. CSIC). Group of Molecular Astrophysics. c/ Serrano 123, 28006 Madrid, Spain.}

\author{Hassan Sabbah}
\affiliation{IRAP, Universit\'e de Toulouse, CNRS, CNES. 9 Av. du Colonel Roche, 31028 Toulouse Cedex 4, France.}

\author{Christine Joblin}
\affiliation{IRAP, Universit\'e de Toulouse, CNRS, CNES. 9 Av. du Colonel Roche, 31028 Toulouse Cedex 4, France.}

\author{Jos\'e Cernicharo}
\affiliation{Instituto de F\'\i{}sica Fundamental (IFF. CSIC). Group of Molecular Astrophysics. c/ Serrano 123, 28006 Madrid, Spain.}

\author{Jos\'e \'Angel Mart\'i{}n-Gago}
\affiliation{Instituto de Ciencia de Materiales de Madrid (ICMM. CSIC). Materials Science Factory. Structure of Nanoscopic Systems Group. c/ Sor Juana In\'es de la Cruz 3, 28049 Cantoblanco, Madrid, Spain.}

\begin{abstract}

Interstellar carbonaceous dust is mainly formed in the innermost regions of circumstellar envelopes around carbon-rich asymptotic giant branch (AGB) stars. In these highly chemically stratified regions, atomic and diatomic carbon, along with acetylene are the most abundant species after H$_2$ and CO. In a previous study, we addressed the chemistry of carbon (C and C$_2$) with H$_2$ showing that acetylene and aliphatic species form efficiently in the dust formation region of carbon-rich AGBs whereas aromatics do not. Still, acetylene is known to be a key ingredient in the formation of linear polyacetylenic chains, benzene and polycyclic aromatic hydrocarbons (PAHs), as shown by previous experiments. However, these experiments have not considered the chemistry of carbon (C and C$_2$) with C$_2$H$_2$.

In this work, by employing a sufficient amount of acetylene, we investigate its gas-phase interaction with atomic and diatomic carbon. We show that the chemistry involved produces linear polyacetylenic chains, benzene and other PAHs, which are observed with high abundances in the early evolutionary phase of planetary nebulae. More importantly, we have found a non-negligible amount of pure and hydrogenated carbon clusters as well as aromatics with aliphatic substitutions, both being a direct consequence of the addition of atomic carbon. The incorporation of alkyl substituents into aromatics can be rationalized by a mechanism involving hydrogen abstraction followed by methyl addition. All the species detected in gas phase are incorporated into the nanometric sized dust analogues, which consist of a complex mixture of sp, sp$^2$ and sp$^3$ hydrocarbons with amorphous morphology.

\end{abstract}


\section{Introduction}

Carbonaceous dust is ubiquitously found in space, from the interstellar medium (ISM) to nova ejecta and circumstellar shells \citep{chiar13}. Astronomical observations in the Mid-Infrared (MIR) range reveal the presence of both aromatic and aliphatic dust components. In the diffuse ISM, the 3.4 $\mu$m absorption band (assigned to the CH vibrational stretching modes of CH$_2$ and CH$_3$ aliphatic moieties), along with the weaker absorption features at 6.8 and 7.3 $\mu$m (which correspond to the CH bending modes of aliphatic groups) have been attributed to hydrogenated amorphous carbon \citep{pendleton02, dartois04}. On the other hand, the so-called aromatic infrared emission bands (AIBs) are widely observed in environments of our Galaxy that are submitted to ultraviolet photons \citep{peeters02, peeters04}. The AIBs, which fall in the spectral range from 3 to 20 $\mu$m (main bands at 3.3, 6.2, 7.7, 8.6 and 11.2 $\mu$m) have generally been assigned to polyaromatic carriers that are small enough to be stochastically heated by the absorption of a single ultraviolet (UV) photon, which constitutes the polycyclic aromatic hydrocarbon (PAH) hypothesis \citep{leger84,  allamandola85, allamandola89, puget89}.  There are also observational evidences for an evolutionary scenario from aliphatics to aromatics calling for very small grains of mixed aliphatic-aromatic composition both in evolved star environments \citep{goto03,kwok11} and in molecular clouds \citep{pilleri15}. Overall, these observations call for a better understanding on the formation of PAHs and of carbon dust in general in astrophysical environments. 

In our Galaxy, carbon-rich (C-rich) asymptotic giant branch (AGB) stars are the major sources of carbonaceous dust \citep{gehrz89}, thus chemical models have been developed to account for the formation of PAHs and hydrogenated carbon clusters in these environments \citep[e.g.][]{pascoli00, cherchneff11}. According to observations, acetylene (C$_2$H$_2$) is one of the most abundant molecules in the inner regions of circumstellar envelopes (CSEs) where dust is formed, presenting a relative abundance to H$_2$ of $8\times10^{-5}$ \citep{fonfria08}, while carbon atoms are also expected to be highly abundant. Chemical equilibrium calculations predict the concentration of carbon atoms to be 1-2 orders of magnitude below that of C$_2$H$_2$ in the region where carbon {molecules are expected to condense into dust \citep{agundez19}.

It is well known that the polymerization of C$_2$H$_2$  gives rise to polyacetylenic chains \citep{cernicharo04}, whose formation mechanism has been recently mapped with ALMA in the outer regions of the C-star envelope IRC\,+10216 \citep{agundez17}. Furthermore, diacetylene (C$_4$H$_2$) has been recently detected in this star through high spectral resolution MIR observations \citep{fonfria18}. Acetylene is also believed to play a crucial role in the formation of benzene and PAH compounds \citep{frenklach89}. The formation of benzene in evolved stars has been rationalized from experiments of C$_2$H$_2$ pyrolysis by the reaction of two propargyl (C$_3$H$_3$) radicals \citep{miller92}, a mechanism that  requires high temperatures and high acetylene concentrations, which are not encountered in the dust nucleation zone of AGB stars. The temperature in the region between 2 and 4 stellar radii is estimated to range from 1500 K to 1000 K whereas a density of acetylene of 10$^7$-10$^5$ molecules/cm$^3$ has been derived by \cite{fonfria08}. This abundance for acetylene is in very good agreement with recent model predictions for for C-rich AGB stars \citep{agundez19}.

Another chemical route for the formation of benzene, considering neutral-neutral reactions, and at much lower temperatures than those needed in the combustion theory, has been derived from acetylene discharges. It proceeds mainly by the addition of C$_2$H$_2$ to the C$_4$H$_3$ radical and to a lesser extent by the cyclization of C$_6$H$_4$ \citep{blecker06_2}. Further growth from benzene to PAHs is assumed to take place through the so-called Hydrogen Abstraction Acetylene Addition (HACA) mechanism \citep{frenklach87, frenklach89, shukla12, yang16}. Recently, a chemical route for the growth of large benzoid-PAHs (comprised only of 6-membered ring structures) has been proposed through the so-called Hydrogen Abstraction Vinylacetylene Addition (HAVA) mechanism \citep{zhao18}. Thus, the complementary HACA/HAVA mechanisms might be needed for the molecular growth of PAHs in circumstellar environments.

Several laboratory experiments have dealt with the formation of stardust analogues from acetylene, either in dusty plasmas \citep{kovacevic05, stefanovic05}, in pyrolysis experiments \citep{jager09, biennier09} or in molecular jets exposed to plasma discharges \citep{contreras13}. However, all of them lack of the addition of atomic carbon and therefore are not suitable to evaluate its impact on the growth of hydrocarbons with implications on, e.g., the formation of hydrocarbons with an odd number of carbon atoms \citep{cernicharo04}. 

Recently, we have shown that sputtering gas aggregation sources (SGAS) are particularly suited for studying dust formation in the CSE of AGBs \citep{martinez19}. In particular, we have demonstrated that the interaction of atomic carbon with molecular hydrogen promotes the formation of non-aromatic molecules, being acetylene one of the main gas phase products. Here, we expand our previous work to investigate the interaction of atomic carbon with acetylene due to its known importance in the formation and growth of hydrocarbons. In order to accelerate the chemistry between carbon and acetylene so as to have access to most of the reaction pathways, as well as to investigate different conditions which would pertain to different regions of the CSE, we have intentionally increased the concentration of acetylene. In addition, our experimental conditions are also of interest for studying the envelopes of C-rich protoplanetary nebulae (PPNe) in which the UV radiation from the central star provokes the photodissociation of acetylene and methane providing a high abundance of carbon initiating a rich chemistry involving C, C$_2$H$_2$, CH$_4$ and other hydrocarbons \citep{cernicharo04}.

We have found that, unlike what is commonly observed for dust analogues prepared in pyrolysis experiments or plasmas, a non-negligible amount of the produced material consists of carbon clusters (both pure and hydrogenated) with an odd number of carbon atoms, being a direct consequence of the interaction of atomic carbon with acetylene. Additionally, aromatics with aliphatic substitutions have been observed. We suggest that the formation of alkyl-substituted aromatics proceeds through the formation of CH$_3$ radicals, which is also a consequence of the addition of atomic carbon.

\section{Experiments}

For the production and analysis of the dust analogues we employed the Stardust machine. The technical details of Stardust can be found elsewhere \citep{martinez18, martinez19}. The particular experimental setup used in this work is schematically depicted in Fig.~\ref{fig:santorof1}. A pictorial view of the CSE region simulated by our experimental setup can be found in \cite{martinez19}. The base pressure of the system is in the 10$^{-10}$ mbar range and the pressure during the production of the analogues at different positions of the machine is also indicated in Fig.~\ref{fig:santorof1}.

Dust analogues were produced with a scaled-up Multiple Ion Cluster Source (MICS) \citep{martinez2012}, a special type of sputtering gas aggregation source (SGAS), working in UHV conditions \citep{haberland91}. Atomic carbon was delivered by magnetron sputtering of a 2-inch graphite target (99.95 \% purity) using Ar as sputtering gas with a flow rate of 150 sccm. The sputtered atoms coalesce inside the aggregation zone (aggregation length, i.e., distance from the magnetron target to the exit nozzle: 374 mm). C$_2$H$_2$ (purity $\geq$ 99.6 \% , diluted in acetone) was injected in the aggregation zone of the MICS through a lateral entrance at constant flow rates of 0,  $4\times10^{-4}$, 0.15 and 1 sccm. The DC magnetron is regulated in current and 0.2 A (corresponding to 100 W) were applied.

Optical emission spectroscopy (OES) was performed inside the aggregation zone (at OES position in Fig. 1) and the light emitted by the sputtering plasma at a distance of around 1 cm from the magnetron target was collected through a fused silica window and a fused silica optical fibre. A 193 mm focal length, motorized Czerny-Turner spectrograph (Andor, model Shamrock SR-193-i-A) equipped with a CCD camera (iDus DU420A-BVF) was employed. Two diffraction gratings with 1200 and 1800 grooves/mm, installed in a movable turret, provide spectral ranges of 300-1200 nm and 200-950 nm, respectively, and nominal spectral resolutions of 0.22 nm and 0.15 nm, respectively (for an input slit width of 20 $\mu$m). The relative spectral efficiencies of the spectroscopic equipment were quantified for both diffraction gratings with a calibrated tungsten lamp.

\begin{figure}
        \centering
        \includegraphics[width=1\linewidth]{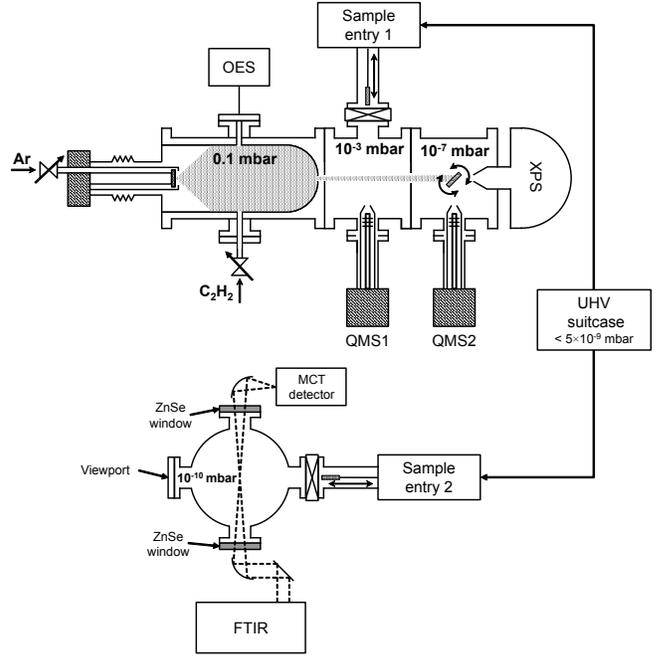}
        \caption{Scheme of the experimental setup used for the analog production and analysis. The pressure at each UHV chamber during the formation of the analogues is indicated. }
        \label{fig:santorof1}
\end{figure}

Mass spectrometry was used for \textit{in-situ} analysis of the gaseous species produced during the production of the dust analogues. This was performed at positions labelled as QMS1 and QMS2 in Fig.~\ref{fig:santorof1}. At QMS1 a quadrupole mass spectrometer (PrismaPlus, Pfeiffer) with a mass range of 0-100 amu and a Faraday cup detector was used. At QMS2, a Pfeiffer HiQuad QMG 700 with QMA 400 mass spectrometer (mass range of 0 to 512 amu) equipped with a CP 400 ion counter preamplifier was used, which enables a higher sensitivity. The pressure in the UHV chamber at position QMS1 during the analogue production precludes the use of a more sensitive Secondary Electron Multiplier (SEM) detector. 
Specific chemical compounds have been tentatively assigned from mass spectra considering the fragmentation pattern by electron impact ionization {available at the NIST database \citep{nist}.

Morphological characterization of the analogues was performed \textit{ex-situ} by Atomic Force Microscopy (AFM) and Transmission Electron Microscopy (TEM). In both cases the deposits were collected through Sample entry 1 of Fig~\ref{fig:santorof1}. For the AFM measurements the dust analogues were collected on SiO$_x$ substrates whereas carbon grids were employed for TEM. The AFM measurements were carried out with Cervantes AFM System equipped with the Dulcinea electronics from Nanotec Electronica S.L. All images were recorded and analyzed using the WSxM software \citep{horcas07}. TEM was performed with a FEI-TITAN X-FEG transmission electron microscope used in scanning mode (STEM) and operated at 300 kV and at 120 kV. The images were acquired using a high-angle annular dark field (HAADF) detector. The microscope was equipped with a monochromator, Gatan Energy Filter Tridiem 866 ERS, a spherical aberration corrector (CEOS) for the electron probe (which allows for an effective spatial resolution of 0.08 nm) and an energy dispersive X-ray detector for EDS analysis.

The composition of the dust analogues was investigated \textit{in-situ} by infrared spectroscopy in transmission geometry. The dust analogues were deposited on KBr substrates and transferred to a separated UHV chamber by means of an UHV suitcase (P $<$  $5\times10^{-9}$ mbar). This protocol ensures that the inspected deposits were not air-contaminated. A VERTEX 70 V (Bruker) instrument was employed and the complete optical path was kept in vacuum. The spectral resolution was set to 2 cm$^{-1}$ and 256 scans were coadded. A liquid nitrogen cooled Mercury Cadmium Telluride (MCT) detector was used. For the analysis of the IR spectra, spectral deconvolution was performed when peaks were not resolved due to overlapping of IR bands.

Finally, the molecular composition of the analogues was investigated \textit{ex-situ} by laser desorption/ionization mass-spectrometry (LDI-MS) employing the AROMA (Aromatic Research of Organics with Molecular Analyzer) setup \citep{sabbah17} using the so-called laser desorption-laser ionization (L2MS) scheme in which desorption and ionization are separated in time and space and performed with two different lasers. The desorption step was performed with the fundamental mode of a Nd:YAG laser ($\lambda$ = 1064 nm) whereas the ionization step was performed employing the fourth harmonic of a Nd:YAG laser ($\lambda$ = 266 nm). This scheme is particularly sensitive to PAHs due to the wavelength of the ionization laser. The ions generated are stored and thermalized in a linear quadrupole ion trap and subsequently monitored by time-of-flight mass spectrometry.

\section{Results}
\subsection{Gas-phase molecules produced during the formation of the dust analogues}

Fig.~\ref{fig:santorof2}a shows the optical emission spectra during the dust analogue formation inside the aggregation zone of the MICS in the spectral regions corresponding to the CH (A-X) band, the C$_2$ Swan band (d${^3}{\Pi}{^g}$-a${^3}{\Pi}{^u}$) and the H$_{\alpha}$ line of the Balmer series for the different C$_2$H$_2$ flow rates employed. When no C$_2$H$_2$ was injected inside the aggregation zone, the molecular C$_2$ Swan band is clearly observed. Since the sputtering of graphite in SGAS produces mainly atomic carbon, we rationalize that C$_2$ is formed via the three body reaction C + C + Ar $\to$ C$_2$ + Ar \citep{martinez19}, which involves Ar atoms from the sputtering gas used.

As C$_2$H$_2$ is injected, the C$_2$ Swan band vanishes and a clear signature of the presence of atomic hydrogen is observed through the appearance of the H$_{\alpha}$ line. This suggests that the excited C$_2$ is consumed to some extent in the reaction C$_2$ + C$_2$H$_2$ $\to$ C$_4$H + H, which is known to be rapid even at low temperatures \citep{canosa07, daugey08}. In addition, as C$_2$H$_2$ is injected, it can directly react with atomic carbon through, e.g., C + C$_2$H$_2$ $\to$ C$_3$H + H and/or C + C$_2$H$_2$ $\to$ C$_3$ + H$_2$ \citep{clary02}, which diminishes the amount of atomic carbon available to form C$_2$ and releases atomic hydrogen as well. Furthermore, emission from the CH radical is detected for the two highest flow rates investigated. We have already shown that this radical can be formed by the interaction of atomic carbon with H$_2$ at very low densities, even with the residual H$_2$ in the UHV chamber \citep{martinez19}. The concurrent detection of the H$_\alpha$ line suggests an excess of atomic hydrogen, which, if it is not participating in further chemical reactions, will recombine into H$_2$ by three-body reactions, thus increasing the overall H$_2$ density. The bimolecular reaction  C + H$_2$ $\to$ CH + H is highly endothermic unless atomic carbon is electronically excited \citep{sato98}. The reaction of C with vibrationally excited H$_2$ could  help in overcoming the barrier energy but under the physical conditions of the Stardust machine we do not expect high abundance of H$_2$ in the v=1, 2  vibrational levels. Therefore, CH is most probably formed by a three body reaction involving Ar (sputtering gas) as third body (C + H + Ar $\to$ CH + Ar), though we cannot completely rule out a contribution from the dissociation of C$_2$H$_2$ in the magnetron plasma. From gas flow calculations, we have estimated that around 2$\%$ of the injected acetylene reaches the magnetron and is therefore susceptible to dissociate by electron impact.
\begin{figure}
        \centering
        \includegraphics[width=1\linewidth]{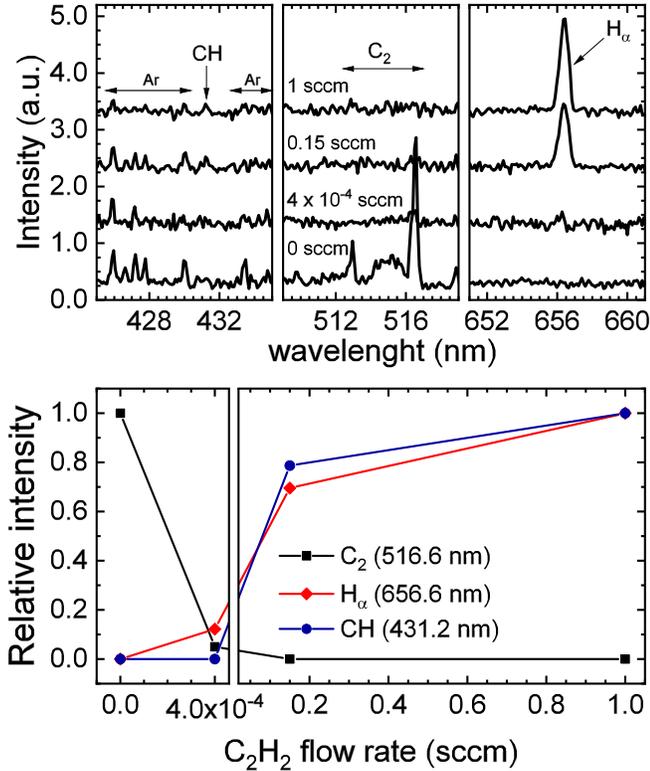}
        \caption{a) Optical emission spectra in the spectral regions of the CH (A-X) band, C$_2$ Swan band (D${^3}{\Pi}{^g}$-a${^3}{\Pi}{^u}$) and the H$_{\alpha}$ line of the Balmer series for the different C$_2$H$_2$ flow rates employed. The flow rates are indicated in the figure. The spectra are vertically shifted for clarity. b) Evolution of the CH, C$_2$ and H$_{\alpha}$ line intensities with the C$_2$H$_2$ flow rate. The wavelength employed for each line is indicated in the figure.}
        \label{fig:santorof2}
\end{figure}

In order to investigate the molecular species that are formed during the dust analogue production, we have performed \textit{in-situ} mass spectrometry at position QMS1 in Fig.~\ref{fig:santorof1}. Once at this position, no further growth of the analogues takes place. When the production of the analogues is carried out in the absence of C$_2$H$_2$, only Ar, which is used as sputtering gas, and the residual gases in the chamber (H$_2$, H$_2$O and CO) were detected (Fig.~\ref{fig:santorof3}a). The peak at m/z = 80 corresponds to the formation of Ar dimers (Ar$_2$) during the gas expansion through the nozzle of the MICS.

On the contrary, when C$_2$H$_2$ is injected we detected diacetylene (C$_4$H$_2$) and for the highest flow rates employed also triacetylene (C$_6$H$_2$) (Fig.~\ref{fig:santorof3}a). The polymerization of acetylene to larger linear polyacetylenic chains is known to occur through the reaction with the ethynil radical (C$_2$H) via C$_2$H + C$_{2n}$H$_2$ $\to$ C$_{2n+2}$H$_2$ + H. More interestingly, for the highest C$_2$H$_2$ flow rate we have clearly detected C$_6$H$_4$ and C$_6$H$_6$ (Fig. ~\ref{fig:santorof3}c). C$_6$H$_4$ has been suggested to be involved in the formation of benzene through a cyclization mechanism \citep{blecker06_2} and the concomitant detection of both species in our experiments seems to support this mechanism.

\begin{figure}
        \centering
        \includegraphics[width=1\linewidth]{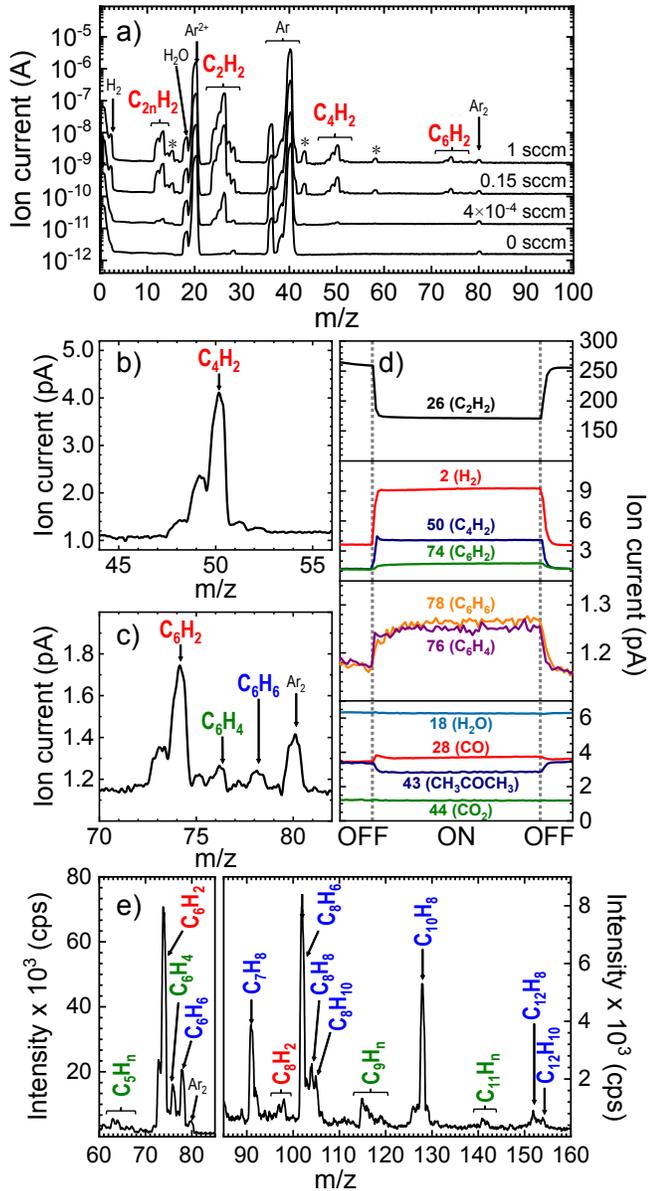}
        \caption{a) \textit{In-situ} mass spectra at position QMS1  in Fig. 1 for the different C$_2$H$_2$ flow rates employed (indicated in the figure). The peaks labelled by a star correspond to the acetone impurity of the C$_2$H$_2$ gas. The curves have been shifted for clarity. b) and c) Expanded regions for the 1 sccm flow rate. d) \textit{In-situ} gas phase detection of m/z = 2, 18, 26, 28, 43, 44, 50, 74, 76 and 78, which are indicated in the figure along with the assignment to the main contributing chemical species for the 1 sccm flow rate. The dashed vertical lines indicate the switching on and off of the magnetron. e) \textit{In-situ} mass spectra for the 1 sccm C$_2$H$_2$ flow rate at position QMS2 in Fig.1. Peaks labeled in red are assigned to polyacetylenic chains, in blue to aromatic molecules and in green to aliphatics. In a) C$_{2n}$H$_2$ label mass peaks associated to the fragmentation of polyacetylenic chains by electron impact in the quadrupole.}
        \label{fig:santorof3}
\end{figure}

As shown in Fig.~\ref{fig:santorof3}d, the formation of all the molecules detected by mass spectrometry is accompanied with the consumption of C$_2$H$_2$ once atomic carbon is released from the magnetron target. In addition, we have observed the production of H$_2$ (m/z = 2). Atomic hydrogen is produced in high amounts by the polymerization reactions of C$_2$H$_2$, thus it is available to participating in further chemical reactions, including its recombination to molecular hydrogen by three body reactions.

The signals corresponding to H$_2$O (m/z = 18) and CO$_2$ (m/z = 44) were stable, indicating that they were not involved in the chemical reactions. However, we observed slight changes in the signal associated to the main fragment of acetone (m/z = 43), which is present as an impurity in the C$_2$H$_2$ (concentration lower than 0.3 \%), along with a slight increase in the amount of CO (m/z = 28). Nevertheless, we have not detected any other gas-phase O-bearing molecule, implying that the acetone impurity in acetylene is only weakly (if any) contributing to the chemistry involved in the formation of the analogues. Moreover, the reaction between C and H$_2$O has a very large barrier of $\simeq$20000 K \citep{mayer67} and that of C with CO$_2$ has been measured at 300 K with a rate below 10$^{-15}$ \citep{husain75}; thus, the formation of O-bearing species is very unlikely in our experiments.

Higher sensitivity gas-phase mass spectrometry was performed at position QMS2 in Fig.~\ref{fig:santorof1} for the highest C$_2$H$_2$ flow rate (Fig.~\ref{fig:santorof3}e). The differences observed in the relative intensities among C$_6$H$_2$, C$_6$H$_4$, C$_6$H$_6$ and Ar$_2$ in Figures 3 (c) and (e) are due to the different sensitivity of the QMS as well as the different pressures at locations QMS1 and QMS2. Concerning polyacetylenic chains, we detected up to C$_8$H$_2$. Furthermore, apart from C$_6$H$_6$, we observed the formation of larger aromatic molecules including naphthalene (C$_{10}$H$_8$), acenaphthylene (C$_{12}$H$_8$) and biphenyl/ethenylnaphthalene (C$_{12}$H$_{10}$). These molecules are important for the growth of larger PAHs through different mechanisms including the well-known HACA mechanism \citep{shukla12}. We have also detected phenylacetylene (C$_8$H$_6$) which is a stable intermediate in the growth of naphthalene from benzene through the HACA mechanism. Interestingly, unlike what is observed in acetylene pyrolysis \citep{shukla12} or acetylene discharges \citep{deschenaux99} we have detected aromatic molecules with aliphatic substitutions (toluene (C$_7$H$_8$), styrene (C$_8$H$_8$) and xylene/ethylbenzene (C$_8$H$_{10}$)). The HACA mechanism does not explain the formation of aromatic compounds with aliphatic substituents. However, the addition of atomic carbon and its interaction with molecular hydrogen (both residual in the UHV system and formed by recombination of the released atomic hydrogen) promotes the formation of alkyl radicals, which open up chemical routes for the formation of alkyl-substituted aromatics (see Sec.~\ref{sec:discussion}).

Finally, we detected a number of hydrogenated carbon clusters (HC-clusters) of aliphatic nature with odd number of carbon atoms from C$_5$ to C$_{11}$, again differing from what is found in acetylene discharges/pyrolysis in which predominantly molecules with even number of carbon atoms are formed (see Fig.~\ref{fig:santorof3}e). The apparent absence of C$_7$ HC-clusters might be due to a blurring of the signal by the peaks associated to toluene and its main fragment (m/z = 91/92). For C$_3$ clusters, it is the Ar signal which precludes its detection. The production of these C-clusters is a clear signature of the interaction of atomic carbon with C$_2$H$_2$. The reaction of C and C$_2$H$_2$ is known to be fast, yielding C$_3$ and/or C$_3$H \citep{liao95,clary02}. Analogous reactions of C with larger polyacetylenic chains are as well an efficient way to produce hydrocarbons with an odd number of carbon atoms in chemical environments where C and C$_2$H$_2$ are abundant \citep{cernicharo04}. In addition, the reaction of C with other large hydrocarbons is also known to occur very fast and without barrier \citep{haider92, liao95, husain97}.

Overall, the results concerning \textit{in-situ} mass spectrometry evidence the formation of three different families of gas-phase chemical compounds, namely, polyacetylenic chains, aromatic species, either pure and with aliphatic substitutions, as well as C- and HC-clusters of aliphatic nature. The identified masses are listed in Table~\ref{table:1} red along with their tentative assignments to chemical compounds \citep{nist}.} A detailed description of the possible formation mechanisms is provided in Sec.~\ref{sec:discussion}.

\begin{table}
\caption{Chemical species detected in the gas phase by mass spectrometry during the production of the dust analogues along with its tentative assignment to specific compounds.}              
\label{table:1}      
\centering                                      
\begin{tabular}{c c c}           
\hline\hline                        
m/z	&		Chemical formula	&			Compound  \\
\hline
\multicolumn{3}{c}{Polyacetilenic chains}  \\

26	&		C$_2$H$_2$				&			acetylene  \\
50	&		C$_4$H$_2$				&			diacetylene  \\
74	&		C$_6$H$_2$				&			triacetylene  \\
98	&		C$_8$H$_2$				&			octatetrayne  \\
\hline
\multicolumn{3}{c}{Aromatics}  \\

78	&		C$_6$H$_6$				&			benzene  \\
91/92	&		C$_7$H$_8$			&			toluene  \\
102	&		C$_8$H$_6$				&			phenylacetylene  \\
104	&		C$_8$H$_8$				&			styrene  \\
106	&		C$_8$H$_{10}$			&			xylene/ethylbenzene  \\
128	&		C$_{10}$H$_8$			&			naphthalene  \\
152	&		C$_{12}$H$_8$			&			acenaphthylene  \\
153/154	&		C$_{12}$H$_{10}$	&			biphenyl/vinylnaphthalene  \\
\hline
\multicolumn{3}{c}{Aliphatics}   \\

62-68	&		C$_5$H$_n$				&			C$_5$-clusters  \\
76	&			C$_6$H$_4$				&			3-hexene-1,5-diyne  \\
115-123	&		C$_9$H$_n$				&			C$_9$-clusters  \\
140-144	&		C$_{11}$H$_n$				&			C$_{11}$-clusters  \\

\hline                                             
\end{tabular}
\end{table}

\subsection{Morphology of the dust analogues}

Apart from the gas phase species, most of the material produced during the formation of the dust analogues consists of particles with diameters in the nanometer range (Fig.~\ref{fig:santorof4}). Due to the low kinetic energy of the particles produced using gas aggregation sources, the particles soft land on the substrate and thus retain the gas phase morphology and structure.

Fig.~\ref{fig:santorof4}a and b show typical AFM images of the dust analogues when no C$_2$H$_2$ was injected in the aggregation zone (pure C particles) and for a C$_2$H$_2$ flow rate of 1 sccm, respectively. In both cases nanoparticles (NPs) are observed with mean NP diameters of 9.3 nm and 7.0 nm for pure C NP and for a C$_2$H$_2$ flow rate of 1 sccm, respectively. However, a broader size distribution is obtained when injecting C$_2$H$_2$ (see Fig.~\ref{fig:santorof4} c and d). 
Moreover, the production rate increased enormously when C$_2$H$_2$ was injected (by a factor around of 350 in terms of collected NPs/$\mu$m$^{2}$s), which evidences the interaction of atomic C with C$_2$H$_2$, resulting in a manifest acceleration of the chemistry.

The TEM analysis of the NPs revealed the formation of amorphous NPs irrespective of the injection of C$_2$H$_2$ (Fig.~\ref{fig:santorof4}e-f). The main difference consists in the shape of the particles. Pure C NPs present a well-defined round shape whereas those formed after C$_2$H$_2$ injection tend to agglomerate. This agglomeration is observed in the AFM images as well, but it is better seen in the TEM images.  Diffusion of the NPs once deposited on the substrates is not expected, thus the agglomerates might be already formed in the gas-phase.

\begin{figure}
        \centering
        \includegraphics[width=1\linewidth]{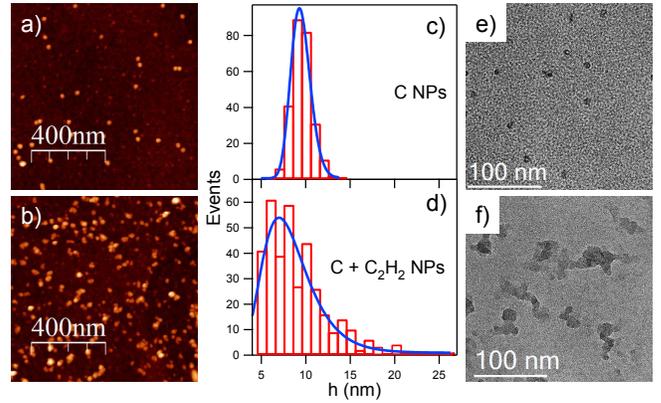}
        \caption{Morphology of the dust analogues without (top) and with (bottom) C$_2$H$_2$ injection (flow rate of 1 sccm). (a,b) AFM images. (c,d) Size distribution extracted from AFM images. The blue lines correspond to the fitting to log-normal distributions. (e-f) TEM images.}
        \label{fig:santorof4}
\end{figure}

\subsection{Composition of the dust analogues}

For the highest C$_2$H$_2$ flow rate (1 sccm), we measured the IR spectrum of the dust analogues (Fig.~\ref{fig:santorof5}). The IR spectrum reveals that the sample consists in a complex mixture of hydrocarbons with sp, sp$^2$ and sp$^3$ carbon hybridizations, thus the gas-phase products detected by mass spectrometry are incorporated into the solid dust analogues. The observed infrared bands are listed in Table~\ref{table:2} along with their tentative assignments.

\begin{deluxetable*}{c c l c}
\label{table:2}
\tablecaption{Vibrational band assignments for the IR bands with optical depth higher than 0.0025.}
\tablewidth{0pt}
\tablehead{
\colhead{$\nu$} & \colhead{$\lambda$} & \colhead{Assignment} &  \colhead{Intensity}\\
\colhead{cm$^{-1}$} & \colhead{$\mu$m}}

\startdata
3322	&		3.01			&			sp alkyne CH str. & sh  \\
3300	&		3.03			&			sp alkyne CH str. & s  \\
3100	&		3.23			&			sp$^2$ arom./olef. CH str. & sh  \\
3085	&		3.24			&			sp$^2$ arom./olef. CH str. & sh  \\
3056	&		3.27			&			sp$^2$ arom./olef. CH str. & m  \\
3029	&		3.30			&			sp$^2$ arom./olef. CH str. & m  \\
2973	&		3.36			&			sp$^3$ asym. CH$_3$ str. & m-s  \\
2925	&		3.42			&			sp$^3$ asym. CH$_2$ str. & s  \\
2870	&		3.48			&			sp$^3$ sym. CH$_3$ str. & sh  \\
2845	&		3.51			&			sp$^3$ sym. CH$_2$ str. & sh  \\
2200	&		4.55			&			sp alkyne CC str. (conjugated alkynes) & w  \\
2105	&		4.75			&			sp alkyne CC str & w  \\
1938	&		5.16			&			Combination of out-of-plane and in-plane CH bend. modes & w  \\
1800-1000	&		5.56-10.00			&			Amorphous carbon ($\pi$-band) & m-s  \\
1711	&		5.84			&			Carbonyl CO str. & w  \\
1686	&		5.93			&			Carbonyl CO str. & w  \\
1630	&		6.13			&			sp$^2$ olef. CC str. & sh  \\
1600	&		6.25			&			sp$^2$ arom. CC str. & m-s  \\
1578	&		6.34			&			sp$^2$ arom. CC str. & sh  \\
1492	&		6.70			&			sp$^2$ arom. CC str. & w  \\
1445	&		6.92			&			sp$^3$ CH$_3$ asym. bend. / sp$^3$ CH$_2$ sci. & m-s  \\
1437	&		6.96			&			sp$^2$ arom. CC str & sh  \\
1379	&		7.25			&			sp$^3$ CH$_3$ sym. bend. & sh  \\
1369	&		7.30			&			sp$^3$ CH$_3$ sym. bend. & w-m  \\
1074	&		9.31			&			sp$^2$ arom. CH in-plane bend. / sp$^2$ olef. CH$_2$ rock. & w\\
1029	&		9.72			&			sp$^2$ arom. CH in-plane bend. & w \\
988	&		10.12			&			sp$^2$ olef. CH wag. & w  \\
968	&		10.33			&			sp$^3$ CH$_3$ rock. & w  \\
910	&		10.99			&			sp$^2$ olef. CH wag. & sh  \\
890	&		11.24			&			sp$^2$ arom. CH out-of-plane bend. (solo) & m  \\
843	&		11.86			&			sp$^2$ arom. CH out-of-plane bend. (duo) & m  \\
776	&		12.89			&			sp$^2$ arom. CH out-of-plane bend. (trio) & sh  \\
756	&		13.23			&			sp$^2$ arom. CH out-of-plane bend. (quartet) & m-s  \\
700	&		14.29			&			sp$^2$ CH out-of-plane bend. & m  \\
640	&		15.63			&			sp$^2$ arom. CCC in-plane bend. & m  \\
\enddata
\tablecomments{arom: aromatics; olef: olefins. The vibrational modes are abbreviated as str: stretching; bend: bending; sci: scissoring; rock: rocking; wag: wagging. Relative intensities are labelled as s: strong; m: medium; w: weak; sh: shoulder.}
\end{deluxetable*}

The presence of sp carbon is evidenced by the absorption bands at 3322 and 3300 cm$^{-1}$ corresponding to the $\equiv$C-H stretching modes, and by the band at 2105 cm$^{-1}$, which corresponds to the -C$\equiv$C- stretching mode of monosubstituted acetylene. In addition, a second -C$\equiv$C- stretching mode is observed at 2200 cm$^{-1}$ which is ascribed to the stretching mode of conjugated triple bonds -C$\equiv$C-C$\equiv$C- \citep{socrates}. This band is also found in dust from acetylene plasmas \citep{stoykov01} but is often too weak to be observed \citep{kovacevic05, stefanovic05}. In addition, it  is not commonly observed in dust analogues from acetylene pyrolysis \citep{biennier09}. The fact that we observe this band might be related to the addition of atomic carbon and the formation of C$_2$, which in the end increases the formation of linear polyacetylenic chains via the C$_4$H radical route (see Sec.~\ref{sec:discussion}).

\begin{figure}
        \centering
        \includegraphics[width=1\linewidth]{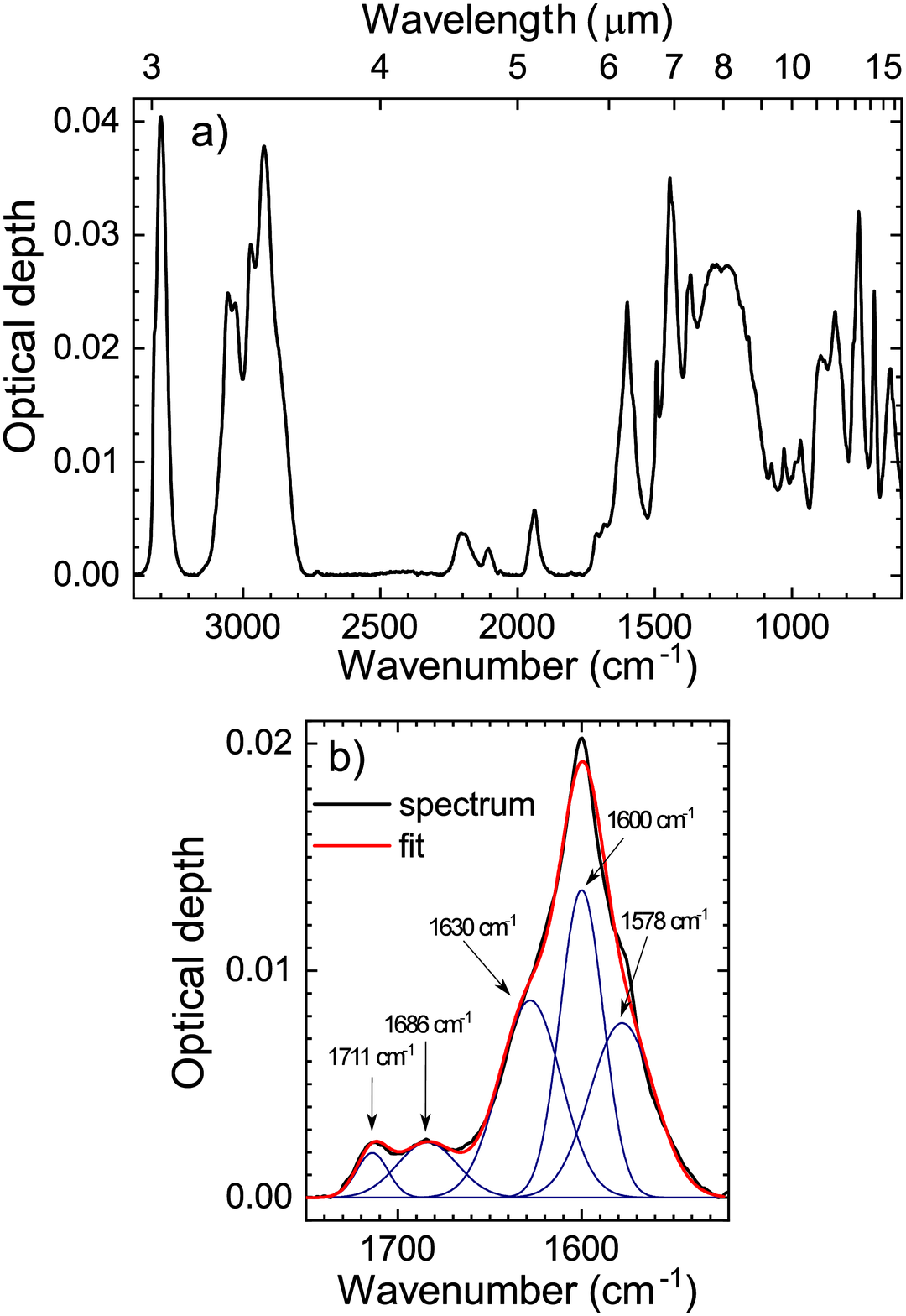}
        \caption{a) IR spectrum of the dust analogue. b) Spectral deconvolution in the range 1500-1750 cm$^{-1}$. For the deconvolution, a baseline has been subtracted to remove the contribution from the very broad amorphous carbon band.}
        \label{fig:santorof5}
\end{figure}

On the other hand, numerous bands indicate the presence of sp$^3$ carbon and aliphatic moieties. The bands at 1369 cm$^{-1}$, 1379 cm$^{-1}$ and 1445 cm$^{-1}$, assigned to the CH$_3$ symmetric bending and CH$_3$ asymmetric  bending / CH$_2$ scissoring modes, respectively (see Table~\ref{table:2}), are a clear signature of alkyl moieties. These bands are accompanied by the bands in the region between 3000 cm$^{-1}$ and 2800 cm$^{-1}$, which correspond to the CH$_2$ and CH$_3$ stretching modes. Moreover, the band at 968 cm$^{-1}$ is ascribed to the CH$_3$ rocking mode.

Concerning sp$^2$ carbon, the bands at 1600 cm$^{-1}$, 1578 cm$^{-1}$, 1492 cm$^{-1}$ and 1437 cm$^{-1}$ reveal the presence of aromatics in the dust analogues. All are assigned to aromatic C=C stretching modes. In the case of alkyl substituted aromatics, the band at 1578 cm$^{-1}$ is usually a shoulder of that at 1600 cm$^{-1}$, as is our case. Moreover, the band at 1492 cm$^{-1}$ is intense for substituted aromatics \citep{socrates}. These results are consistent with the detection of aromatics with aliphatic substituents by mass spectrometry and reveal that the aromatic compounds incorporated into the dust analogues are not only pure but present aliphatic substituents as well. We note that a similar result has been recently obtained by \cite{gavilan20} during the formation of dust analogues using small PAHs as precursors in a molecular jet exposed to an electrical discharge.

\begin{figure}
        \centering
        \includegraphics[width=1\linewidth]{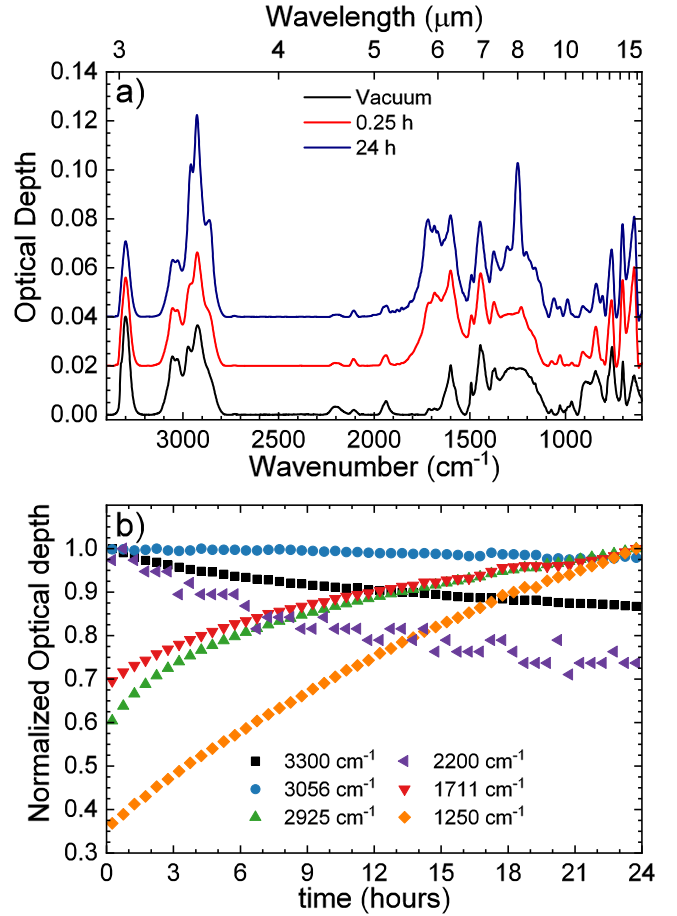}
        \caption{a) IR spectra of the dust analogues in UHV and after air exposure for 15 min and 24 hours. The spectra have been vertically shifted for clarity. b) Temporal evolution of the normalized optical depth for some selected IR bands after exposing to air the dust analogue.}
        \label{fig:santorof6}
\end{figure}

Other bands related to the aromatic compounds incorporated into the dust analogues are the CH out-of-plane bending modes. The different bands observed at 890 cm$^{-1}$, 843 cm$^{-1}$, 776 cm$^{-1}$ and 756 cm$^{-1}$ are related to different number of peripheral adjacent H atoms (solo, duo, trio and quartet, respectively) \citep{hony01, carpentier12} and are characteristic of PAHs though insensitive to PAH substitution. However, the band at 700 cm$^{-1}$ can be assigned to CH out-of-plane bending mode of substituted benzenes \citep{socrates}. Finally, the band at 640 cm$^{-1}$ is attributed to the CCC in plane bending of aromatic compounds \citep{gadallah12} and the combination band at 1938 cm$^{-1}$ is ascribed to the combination of out-of-plane and in-plane CH bending modes (combination and overtone bands in the spectral region 2000-1700 cm$^{-1}$ are characteristic of aromatic compounds).

Apart from aromatic compounds, there are also bands related to olefinic sp$^2$ carbon. The sp$^2$ CH stretching bands in the region 3100-3000 cm$^{-1}$ are due to both aromatic and olefinic sp$^2$ carbon, but the shoulder at 1630 cm$^{-1}$ is unambiguously assigned to the C=C stretching modes of olefinic compounds (Fig.~\ref{fig:santorof5}b) \citep{socrates,gavilan17} . The presence of olefinic moieties is further confirmed by the characteristic wagging modes of olefins at 988 cm$^{-1}$ and 910 cm$^{-1}$ \citep{socrates}. The band at 1074 cm$^{-1}$ might be due to the CH$_2$ rocking mode characteristic of vinyl groups, in accordance with the observation of styrene (and maybe also vinylnaphthalene) by mass spectrometry, though the aromatic in-plane CH bending modes overlap in this region.

The broad band that appears in the 1800-1000 cm$^{-1}$ region is characteristic of amorphous carbon (aC) and its intensity is related to the relative amount of sp$^2$ to sp$^3$ carbon hydridization \citep{ferrari03, rodil05}.  Its presence in the IR spectrum is consistent with the amorphous morphology observed by TEM and suggests that part of the dust analogues is made of aC material.

Finally, weak C=O carbonyl stretching bands are observed which we consider as related to the acetone impurity in the C$_2$H$_2$ bottle since the base pressure of the Stardust machine is 10$^{-10}$ mbar and the sample was transferred to the IR UHV chamber by means of an UHV suitcase at a pressure lower than $5\times 10^{-9}$ mbar.

In addition, we have tested the stability of the analogues after air exposure and we have observed that they present a high reactivity (Fig.~\ref{fig:santorof6}), whereas we did not observe any significant change in the IR spectrum of the analogues after 24 hours in UHV. Carbonaceous materials produced from unsaturated precursors are known to react swiftly with oxygen, leading to the formation of C-O-C moieties. Mainly, we have observed the incorporation of oxygen through the increase of the carbonyl C=O stretching modes and through the appearance and growth of a band at 1250 cm$^{-1}$, assigned to the C-O-C stretching mode. In addition we have observed a reduction in the sp carbon content, which is evident by the decrease of the sp CH stretching modes (3300 cm$^{-1}$) and of the -C$\equiv$C- stretching mode of conjugated  alkynes (2200 cm$^{-1}$). An increase in sp$^3$ aliphatic moieties was also observed whereas the sp$^2$ CH stretching modes remain unaltered (see e.g. the normalized optical depth (OD) for the bands at 2925 cm$^{-1}$ and at 3056 cm$^{-1}$ in Fig.~\ref{fig:santorof6}b). Further changes in the 1000-600 cm$^{-1}$ region can also be observed, though difficult to interpret. Anyhow, these results evidence the extreme importance of both working in ultra-clean environments (such as UHV conditions) and of \textit{in-situ} characterizing the dust analogues produced from acetylene.

Albeit the tested chemical reactivity of the dust analogues towards atmospheric air, we have also performed \textit{ex-situ} laser desorption/ionization mass-spectrometry (LDI-MS) of the dust analogue produced with a C$_2$H$_2$ flow rate of 1 sccm (Fig.~\ref{fig:santorof7}), using the AROMA setup \citep{sabbah17}.The results are not quantitative in terms of the relative amount of the observed compounds because of the known chemical evolution of the analogues after air exposure, but they provide additional information on the chemical composition.

\begin{figure}
        \centering
        \includegraphics[width=1\linewidth]{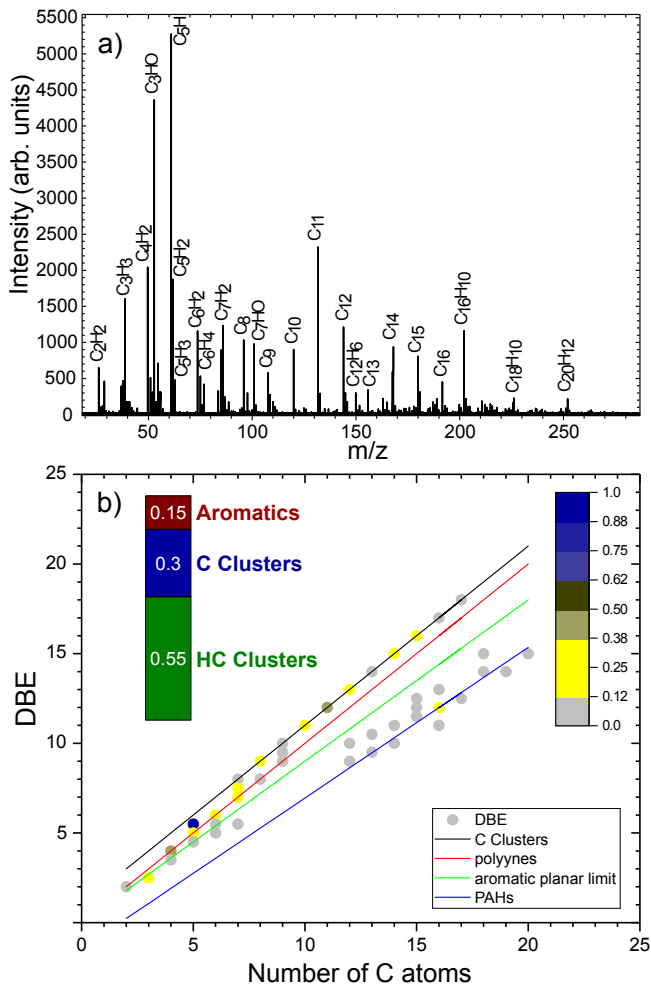}
        \caption{a) \textit{Ex-situ} laser desorption/ionization mass-spectrometry (LDI-MS) of the dust analogue using the AROMA setup \citep{sabbah17}. b) Double Bond Equivalent (DBE) analysis. The stacked bar graph in the inset summarizes the composition of each molecular family.}
        \label{fig:santorof7}
\end{figure}

Linear polyacetylenic chains (in cationic form) were detected up to C$_8$H$_2$, which is somehow surprising due to the high volatility of these compounds and indicates that these species are trapped inside the dust particles and released by laser desorption. Other volatile species detected by \textit{in-situ} mass spectrometry, such as benzene or toluene, were not detected by LDI-MS despite the high sensitivity of the technique to aromatics \citep{sabbah17}. C$_6$H$_4$ could be identified, which was also detected by \textit{in-situ} mass spectrometry and is an important molecule for the formation of benzene. Moreover, C$_2$H$_2^+$ and C$_3$H$_3^+$ are believed to be produced during the laser desorption stage from the common C$_3$H$_2$ precursor \citep[cf. supplementary material in][]{martinez19}.

On the other hand, C- and HC-clusters with carbon atoms from 5 to 20 dominate the mass spectrum, being C$_5$H and C$_{11}$ particularly abundant. Relatively large aromatics are also present in the dust analogue from C$_{12}$H$_8$ (acenaphthylene) to C$_{20}$H$_{12}$. Finally, the peaks assigned to C$_3$HO and C$_7$HO evidence oxidation of the sample likely due to air exposure. The results from LDI-MS are summarized in Fig.~\ref{fig:santorof7}b by using a Double Bond Equivalent (DBE) analysis. Briefly, in the case of hydrocarbons, DBE can be defined as the sum of the number of double bonds involving carbon plus the number of rings for a particular species. A description of the DBE analysis can be found in, e.g,  \cite{sabbah17}, \cite{gavilan20} and in the Supplementary Information of \cite{martinez19}. Three molecular families, namely aromatics, HC- and C-clusters, were considered and their relative intensities in terms of summed peak intensities are shown.

\section{Discussion} \label{sec:discussion}
\subsection{Formation of polyacetylenic chains (C$_{2n}$H$_2$)}

Linear polyacetylenic chains, C$_{2n}$H$_2$, are observed in the outer layers around C-rich stars and their formation involves the ethynyl radical (C$_2$H), which is formed by the photodissociation of C$_2$H$_2$ in the outer layers of the CSE and triggers the chemistry of linear polyacetylenic chains \citep{cernicharo04,agundez17}. Once the C$_2$H radical is formed, the formation of polyacetylenes proceeds mainly through reactions of the type

\medskip
\centerline{C$_2$H + C$_{2n}$H$_2$ $\to$ C$_{2n+2}$H$_2$ + H				(R1)}
\medskip
which are assumed to occur fast at low temperatures (see scheme in Fig.~\ref{fig:santoros2}) \citep{chastaing98,agundez17}. A second important chemical route for the formation of triacetylene and larger polyacetylenic chains involves the radical C$_4$H through reactions of the type \citep{berteloite10,agundez17}

\medskip
\centerline{C$_4$H + C$_{2n}$H$_2$ $\to$ C$_{2n+4}$H$_2$ +H				(R2).}
\medskip

Besides the photodissociation of diacetylene, the C$_4$H radical can be efficiently formed by the interaction with dicarbon through the bimolecular reaction

\medskip
\centerline{C$_2$ + C$_2$H$_2$ $\to$ C$_4$H +H 			(R3).}
\medskip

\begin{figure}
        \centering
        \includegraphics[width=1\linewidth]{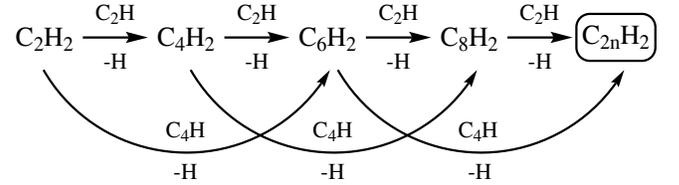}
        \caption{Reaction pathway for the formation of linear polyacetylenic chains.}
        \label{fig:santoros2}
\end{figure}

In CSEs around C-rich AGB stars, the interaction of C$_2$H$_2$ with energetic radiation can lead to dissociation but also to ionization. Taking into account that ion-neutral reactions are usually faster than neutral-neutral reactions (e.g., the rate constant of C$_2$H$_{2}^{+}$ + C$_2$H$_2$ $\to$ C$_4$H$_{2}^{+}$ + H$_2$ is 2.5-5 times higher than that of C$_2$H + C$_2$H$_2$ $\to$ C$_4$H$_2$ + H; \citealt{anicich93,chastaing98}), chemical reactions involving ions may also play a role in the synthesis of polyacetylenic chains. The polymerization of C$_2$H$_2$ in the external layers is dominated by neutral-neutral reactions involving C$_2$H, rather than by ion-molecule reactions involving C$_2$H$_2^+$, because photodissociation of C$_2$H$_2$ by the interstellar UV field is 5-10 times faster than photoionization \citep{heays17,agundez18}. However, in the inner clumps of CSEs, where dust acts as an efficient shielding against UV photons but cosmic rays can penetrate, some C$_2$H$_2$ may be chemically processed due to cosmic-ray induced ionization and further ion-molecule reactions. The C-star envelope IRC\,+10216 is known to present a clumpy structure \citep{cernicharo15} and, in fact, cation-neutral reactions have been suggested to play a role for the formation of, e.g., CH$_3$CN, SiH$_3$CN and CH$_3$SiH$_3$ \citep{agundez08,cernicharo17}. However, the measured abundances of ions (cations and anions) is very low (X$\le$10$^{-10}$), precluding an important role in the chemistry of the most abundant carbon-bearing species (note that the anions of all C$_n$H radicals have been detected in the external and cold layers of the envelope but not in the inner warm regions).

In our experiments, as mentioned in the previous section, we estimate that around 2$\%$ of the injected acetylene reaches the magnetron and can therefore be dissociated into C$_2$H by electron impact (see Fig.~\ref{fig:santoros1} for a scheme on the formation mechanism of radicals from atomic carbon and acetylene in SGAS). The interaction of C$_2$ with H$_2$ (either residual in the chamber as well as produced by recombination of the atomic hydrogen in excess) contributes as well to the formation of the C$_2$H radical via the bimolecular reaction C$_2$ + H$_2$ $\to$ C$_2$H + H \citep{cernicharo04, martinez19}. In addition, and due to the higher cross section for electron impact ionization over electron impact dissociation \citep{mao08}, chemical reactions involving cations such as C$_2$H$_{2}^{+}$ and C$_2$H$^{+}$ might take place. However, the electron temperature of the plasma in magnetron sputtering discharges and in SGAS sources is lower than 10 eV \citep{{ivanov92, kousal17}}, which favours electron impact dissociation over direct ionization or dissociative ionization events. This is different to the case of dusty plasmas in which apart from the neutral dissociative production of C$_2$H, there is a higher probability for the formation of C$_2$H$_{2}^{+}$ (from C$_2$H$_2$) and C$_2$H$^{+}$ (from C$_2$H) by electron-knocking off and dissociative ionization processes, respectively. Thus, the chemistry of the formation of nanoparticles in C$_2$H$_2$ dusty plasmas presents a non-negligible ion-neutral contribution \citep{jimenez19}. In addition, anionic polymeryzation routes can also be relevant in acetylene plasmas \citep{deschenaux99, blecker06, jimenez19}  due to the trapping of negative ions in the plasma potential, despite  the electron attachment probability being several orders of magnitude lower than those for the electron impact dissociation/ionization. However, the chemistry involving anions is very unlikely in SGAS, in which it has been shown that the chemistry proceeds predominantly by neutral-neutral interactions \citep{haberland91, martinez19} .

\begin{figure}
        \centering
        \includegraphics[width=1\linewidth]{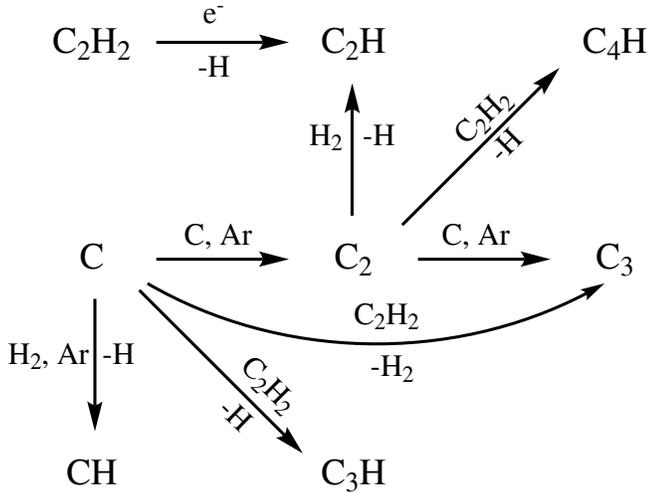}
        \caption{Formation mechanism of radicals from atomic carbon and acetylene in SGAS. Molecular hydrogen is both residual in the UHV system and formed by recombination of the released atomic hydrogen.}
        \label{fig:santoros1}
\end{figure}

On the other hand, the results from OES suggest that  the excited C$_2$ is consumed to some extent in the reaction (R3) to produce the C$_4$H radical. Therefore, during the production of the dust analogues the growth of polyacetylenic chains involves mainly the C$_2$H and C$_4$H radicals (Fig.~\ref{fig:santoros2}). Moreover, the incorporation of conjugated polyalkynes to the dust analogues in higher amounts than those observed in acetylene discharges, as probed by IR spectroscopy, supports a significant contribution of the C$_4$H chemical route, which is initiated by the addition of atomic carbon and subsequent formation of C$_2$ in substantial amounts.

All the above mentioned reactions leading to the growth of polyacetylenic chains release either atomic or molecular hydrogen, which is consistent with the observed increase in the peak at m/z = 2 (Fig.~\ref{fig:santorof3}c) and with the increase of the H$_\alpha$ line in the optical emission spectra (Fig.~\ref{fig:santorof2}a), being therefore available in high amounts to participating in further chemical reactions. We assume that the atomic hydrogen that is not consumed in chemical reactions is recombined into H$_2$ by three body reactions involving both Ar and the solid dust analogues.

\subsection{Formation of PAHs}

Benzene formation is considered as the bottleneck for the growth of PAHs, which have been proposed as the carriers of the AIBs and the seeds of carbonaceous cosmic dust. Benzene has been detected in protoplanetary nebulae \citep{cernicharo01, malek11} but not yet in the CSE of C-rich AGB stars, which constitute the main factories of cosmic dust. In addition, no specific PAH structure has been identified from observations. In particular, the apolar (or weakly polar) character of these molecules along with large rotational partition functions, hinder its detection by rotational emission spectroscopy \citep[e.g.][]{joblin09}. Recently, benzonitrile has been detected in the ISM \citep{mcguire18}, whose formation might be linked to cyanopolyynes in the same manner as benzene is believed to be related to polyynes.

Despite the lack of successful detection of benzene or PAHs in CSEs around C-rich AGB stars, a bottom-up chemical pathway, derived from combustion experiments, has been proposed for the formation of  benzene and PAHs in the warm inner regions of C-rich CSEs \citep{cherchneff92,cherchneff12}. In this scheme, the key reaction is the direct recombination of two propargyl radicals to form both cyclic and linear C$_6$H$_6$ as well as the phenyl radical (C$_6$H$_5$)

\medskip
\centerline{C$_3$H$_3$ + C$_3$H$_3$ $\to$ C$_6$H$_6$				 (R4a)}
\medskip
\centerline{C$_3$H$_3$ + C$_3$H$_3$ $\to$ C$_6$H$_5$ + H 				(R4b)}
\medskip

being reaction (R4a) more efficient than (R4b). A different chemical scheme involving hydrocarbon ions has been proposed to explain the formation of benzene in the protoplanetary nebula CRL 618 \citep{woods02}, where C$_6$H$_6$ has been detected. In this scheme, the synthesis is initiated with the proton transfer to C$_2$H$_2$ to form C$_2$H$_3^+$, followed by successive reactions with C$_2$H$_2$ to form cyclic C$_6$H$_5^+$, and ended with the reaction of C$_6$H$_5^+$ with H$_2$ to form C$_6$H$_7^+$, which yield benzene upon dissociative recombination with electrons.

Apart from a top-down etching process \citep{merino14} not considered here, another possible bottom-up mechanism for the formation of benzene involving neutral-neutral reactions has been derived from plasma discharges \citep{blecker06_2}. In this case, C$_2$H, which is also crucial to the formation of polyacetylenic chains, initiates the chemistry mainly by the termolecular association reaction 

\medskip
\centerline{C$_2$H + C$_2$H$_2$ + M  $\to$ C$_4$H$_3$ + M				 (R5).}
\medskip

where M stands for a third-body.

In our case, due to the temperatures involved in the sputtering process in SGAS (in the order of 500 K) we rule out the formation of benzene via propargyl radicals (note that the C$_3$H$_3^+$ observed by LDI-MS is formed during the laser desorption step by protonation of C$_3$H$_2$). In fact, the concurrent detection of C$_6$H$_4$ and C$_6$H$_6$ by \textit{in-situ} mass spectrometry points towards a similar neutral-neutral route as that proposed for acetylene discharges, involving C$_2$H and the formation of the butadienyl radical (C$_4$H$_3$). Moreover, the formation of C$_4$H$_3$ is less efficient than the formation of C$_4$H$_2$, which explains the observation of C$_6$H$_6$ only at the higher C$_2$H$_2$ flow rates employed in our experiments. Once C$_4$H$_3$ is formed, the main reactions with C$_2$H$_2$ produce C$_6$H$_5$ and C$_6$H$_4$ (Fig.~\ref{fig:santoros3}), which account for about 75 $\%$ and 25$\%$ of the reaction mechanism towards C$_6$H$_6$, respectively \citep{blecker06_2}. In our case, the detection of C$_6$H$_5$ is hindered due to the location of the mass spectrometers, which only allows stable species to be detected.

Once C$_6$H$_6$ is formed, the growth towards PAHs has been considered in astrophysical models to proceed  by the so-called HACA mechanism in which acetylene is the main species propagating the growth of aromatic rings (Fig.~\ref{fig:santoros3}). In addition, the HAVA mechanism \citep{zhao18} as well as fusion of PAHs \citep{shukla10} have been proposed for the growth of large benzoid PAHs. However, none of them explains the formation of aromatics with aliphatic substituents (except those of acetylenic nature), such as those observed in our experiments, both in the gas-phase (by \textit{in-situ} mass spectrometry) and in the solid dust analogues (by \textit{in-situ} IR spectroscopy).

\begin{figure}
        \centering
        \includegraphics[width=1\linewidth]{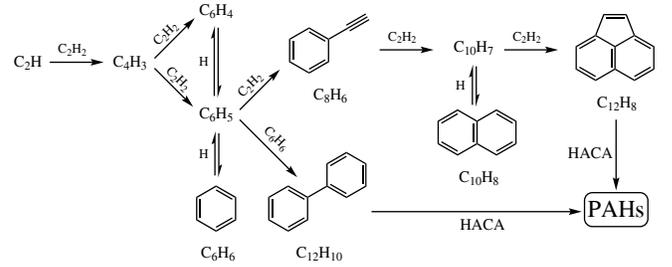}
        \caption{Reaction pathway for the formation of PAHs.}
        \label{fig:santoros3}
\end{figure}

\subsection{Formation of alkyl-substituted aromatics}

Alkyl substituted aromatics have been proposed to be present in space based on the analysis of the 3.3/3.4\,$\mu$m band intensity ratio \citep{joblin96}. \cite{li12} estimated an upper limit of around 15 $\%$ for the aliphatic fraction in PAHs derived from the intensities of the 3.4 $\mu$m and 6.85 $\mu$m emission features. The formation of alkyl substituted aromatics is expected to involve the reactivity of the methyl radical (CH$_3$) with aromatic units and/or aromatic radicals \citep{shukla10b}.

In CSE around C-rich AGB stars, CH$_4$ has been detected \citep{keady93} and it is likely that methyl radicals are present, although they have not been detected and their abundance is unknown. Chemical equilibrium predicts a relatively low abundance of $10^{-9}$ relative to H$_2$ in the stellar atmosphere  \citep{agundez19}, and unfortunately chemical kinetics models of the inner wind (e.g., \citealt{cherchneff12}) have not provided abundance estimates for CH$_3$. A bottom-up synthesis of CH$_3$ must involve the methylidyne radical (CH), which might arise from the interaction of atomic carbon with molecular hydrogen via C + H$_2$ $\to$ CH + H. However, this reaction is endothermic by about 100 kJ/mol and necessitates temperatures in the range of several thousand Kelvin to be efficient. Thus, shock-wave induced chemistry would be needed to efficiently form the methylidyne radical from a bottom-up approach. Once CH radicals are formed, the interaction with H$_2$ through the following reaction

\medskip
\centerline{CH + H$_2$ + M $\to$ CH$_3$ + M				 (R6)}
\medskip

where M stands for a third-body, can lead to CH$_3$. If this is the case, cosmic dust might be needed as third body to efficiently produce CH$_3$. On the other hand, CH$_3$ radicals can be formed in the outer layers of the CSE by photodissociation of CH$_4$ in the same way as C$_2$H radicals are formed from C$_2$H$_2$.

Methylated species have been detected in the C-star AGB IRC+10216 \citep{cernicharo00,agundez15,cernicharo17}. However, their low abundances suggest that they play a minor role in the chemistry of the envelope. A different situation has been found in the CRL618 PPN, where methylpolyynes (CH$_3$C$_2$H and CH$_3$C$_4$H) have been detected with abundances much larger than those observed in IRC+10216 \citep{cernicharo01_b}

Aromatic compounds with aliphatic moieties are neither produced in acetylene pyrolysis \citep{shukla12} nor detected (or only very marginally) as neutral species in the gas phase of acetylene discharges \citep{deschenaux99, consoli08}. However, this is drastically different when other hydrocarbons are used, particularly in ethylene (C$_2$H$_4$) pyrolysis and ethylene plasmas \citep{jager06, gillon14}. This is related to the formation of alkyl radicals which enter into the chemistry, thus promoting the formation of alkyl moieties. \cite{shukla10b} demonstrated that pyrolysis of aromatics in the presence of CH$_3$ radicals can lead to aromatic growth through a Methyl-Addition Cyclization process and\cite{gavilan20} have recently proposed a route for PAH growth at low temperatures (150-200 K) by the addition of alkyl radicals. 

In our case, the CH radical, which we observe by OES, is formed by the reaction 

\medskip
\centerline{C + H + Ar/dust $\to$ CH + Ar/dust				 (R7)}
\medskip

involving Ar and the solid dust analogues as third-body.

Once the CH radical is formed, subsequent three body associations with atomic H would lead to CH$_3$. This chemical route continues to the formation of CH$_4$, but the acetone impurity in the C$_2$H$_2$ bottle prevents its detection by \textit{in-situ} mass spectrometry. Nevertheless, we have recently demonstrated the formation of CH$_4$ and C$_2$H$_4$ by the interaction of atomic C and H$_2$ even at very low H$_2$ concentrations \citep{martinez19}.

The formation of toluene (C$_7$H$_8$) proceeds then by the reaction of phenyl (C$_6$H$_5$) and methyl (CH$_3$) radicals through

\medskip
\centerline{C$_6$H$_5$ + CH$_3$ $\to$ C$_7$H$_8$ + H				 (R8)}
\medskip

being the phenyl radical formed as shown in Fig.~\ref{fig:santoros3}.

The growth towards longer alkyl-chain substitutions to form e.g. ethylbenzene (C$_8$H$_{10}$) proceeds through hydrogen abstraction and methyl addition . This process is not limited to the phenyl radical but applies to any other aromatic radical formed during the growth of PAHs through the HACA mechanism (e.g., naphtyl: C$_{10}$H$_7$, acenaphtyl: C$_{12}$H$_7$, …) and provides a route for the growth of alkyl substituted PAHs (Fig.~\ref{fig:santoros4}), closely related to the Methyl-Addition Cyclization mechanism \citep{shukla10b}.

\begin{figure}
        \centering
        \includegraphics[width=1\linewidth]{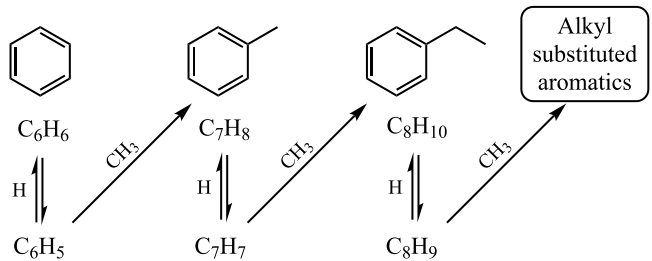}
        \caption{Reaction pathway for the formation of alkyl substituted aromatics through hydrogen abstraction and subsequent methyl addition.}
        \label{fig:santoros4}
\end{figure}

On the other hand, the likely detection of styrene (C$_8$H$_8$) (maybe vinylnaphthalene (C$_{12}$H$_{10}$) as well) by \textit{in-situ} mass spectrometry along with the identification of olefins by IR spectroscopy (including a  band likely related to vinyl moieties) points towards the production of the vinyl radical (C$_2$H$_3$). Bimolecular reactions involving C, C$_2$H$_2$ and H$_2$ to form C$_2$H$_3$ are highly endothermic for all the possible chemical routes. The same holds for the addition of C$_2$H$_3$ to PAHs and aromatic radicals. Therefore, we assume that the formation of aromatics with vinyl substitutions proceeds by three body reactions involving both Ar and the solid dust analogues.

\subsection{Formation of C-clusters with odd C atoms}

Molecules and carbon clusters with odd number of carbon atoms have been detected in the circumstellar environments of C-rich stars \citep{hinkle88, bernath89} as well as in interstellar clouds \citep{maier01}. 
However, the growth of carbon clusters from C and C$_2$ at the temperatures found in the CSE is still not well characterized since most chemical reactions involving small carbon clusters are still unknown. Up to date only the C$_2$+C$_2$ reaction has been measured at high temperature \citep{kruse97}, which has been shown to present a high reaction rate. Therefore, although it potentially contributes to the chemistry, much is yet to be understood on the role of the C$_n$+C$_m$ reactions in the formation of large carbon clusters. An alternative and efficient chemical route is provided by the interaction of atomic carbon with C$_2$H$_2$ through the reactions \citep{cernicharo04}

\medskip
\centerline{C + C$_2$H$_2$ $\to$ C$_3$H + H				 (R9a)}
\medskip
\centerline{C + C$_2$H$_2$ $\to$ C$_3$ + H$_2$				 (R9b).}
\medskip

Additionally, C$_2$H plays again an important role, as in the formation of polyacetylenic chains and benzene, as well as C$_4$H by the interaction with atomic carbon via 

\medskip
\centerline{C + C$_2$H $\to$ C$_3$ + H				 (R10a)}
\medskip
\centerline{C + C$_4$H $\to$ C$_5$ + H				 (R10b).}
\medskip

Once C$_3$ and C$_5$ are formed by (R9) and (R10), the growth towards larger molecules with odd number of carbon atoms proceeds through the following reactions with acetylene

\medskip
\centerline{C$_{2n+1}$ + C$_2$H$_2$ $\to$ C$_{2n+3}$H + H				 (R11a)}
\medskip
\centerline{C$_{2n+1}$ + C$_2$H$_2$ $\to$ C$_{2n+3}$ + H$_2$				 (R11b)}
\medskip

as well as by the following interactions with C$_2$H and C$_4$H

\medskip
\centerline{C$_{2n+1}$ + C$_2$H $\to$ C$_{2n+3}$ + H				 (R12a)}
\medskip
\centerline{C$_{2n+1}$ + C$_4$H $\to$ C$_{2n+5}$ + H					 (R12b).}
\medskip

Therefore, the growth of C-clusters with an odd number of carbon atoms proceeds through two different chemical routes: an acetylenic route (R11) and a radical route (R12). In fact, these two chemical pathways do not apply only to C$_{2n+1}$-clusters but also to C$_{2n}$-clusters, providing a chemical way to the formation of hydrogenated C-clusters.

The above-mentioned mechanisms necessitate atomic carbon to initiate the chemistry, which is not present in acetylene combustion experiments or in acetylene discharges. In fact, in particle-forming acetylene discharges the dominant species are by far hydrocarbon molecules with even number of carbon atoms \citep{deschenaux99, consoli08, kovacevic03, benedikt10}. On the contrary, the supply of atomic carbon in our case opens up both the acetylenic and the C$_2$H/C$_4$H radical routes described, resembling more closely the mechanism operating in the circumstellar environments of C-rich stars. Figure ~\ref{fig:santoros5} summarizes the chemical network for the growth of C-clusters. Three body reactions involving Ar, which cannot be excluded in our experiments, are also shown.

\begin{figure}
        \centering
        \includegraphics[width=1\linewidth]{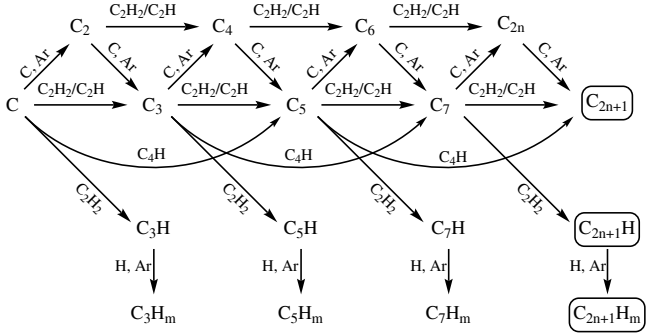}
        \caption{Reaction pathway for the formation of C$_{2n+1}$H$_{m}$ clusters.}
        \label{fig:santoros5}
\end{figure}

\section{Conclusions}

In this article we present a thorough study of the gas-phase chemistry of acetylene in the presence of C and C$_2$ which can pertain to the outer layers of C-richAGB stars and protpplanetary nebulae. We have formed and studied \textit{in-situ} cosmic dust analogues employing atomic carbon and acetylene as precursors, using a a well-controlled ultra-clean experimental set-up. Apart from the formation of linear polyacetylenic chains, we have observed the formation of polyaromatic hydrocarbons (PAHs). More importantly, aromatics with aliphatic substitutions as well as pure and hydrogenated carbon clusters were produced as a direct consequence of the addition of atomic carbon. The growth of carbon clusters proceeds from the interaction of atomic carbon with acetylene whereas the formation of alkyl-substituted aromatics is rationalized by the formation of methyl radicals and subsequent incorporation into aromatics by a hydrogen abstraction-methyl addition mechanism. In addition, the observed gas-phase species incorporate into the dust analogues, which consist in a complex mixture of sp, sp$^2$ and sp$^3$ hydrocarbons with an amorphous morphology. Our results are of particular interest for unveiling chemical routes leading to formation of acetylene-based molecular species in the external layers of AGB stars and in protoplanetary nebulae and foster the search for alkyl-substituted aromatics in these environments. 

\acknowledgments

We thank the European Research Council for funding support under Synergy Grant ERC-2013-SyG, G.A. 610256 (NANOCOSMOS). Also, partial support from the Spanish Research Agency (AEI) through grants MAT2017-85089-c2-1R, FIS2016-77578-R, FIS2016-77726-C3-1-P, AYA2016-75066-C2-1-P, and RyC-2014-16277 is acknowledged. Support from the FotoArt-CM Project (P2018/NMT 4367) through the Program of R$\&$D activities between research groups in Technologies 2013, co-financed by European Structural Funds, is also acknowledged. G.T-C. acknowledges funding from the Comunidad Aut\'onoma de Madrid (PEJD-2018-PRE/IND-9029).

\bibliography{Santoro_C_C2H2}{}
\bibliographystyle{aasjournal}

\end{document}